\documentclass[prb,twocolumn,superscriptaddress,preprintnumbers,a4paper,amsmath,amssymb,showpacs,floatfix]{revtex4-1}
\usepackage{graphicx}
\usepackage{graphics}
\usepackage{bm}
\usepackage{dcolumn}

\newcommand{\urxrs}{U(Ru$_{0.92}$Rh$_{0.08}$)$_{2}$Si$_{2}$}
\newcommand{\urs}{URu$_{2}$Si$_{2}$}

\begin{document}

\title{Field-induced phases in a heavy-fermion \urxrs~single crystal}

\author{K.~Proke\v{s}}
\email{prokes@helmholtz-berlin.de}
\affiliation{Helmholtz-Zentrum
Berlin f\"{u}r Materialien und Energie, Hahn-Meitner Platz 1, 14109 Berlin, Germany}

\author{T. F\"orster}
\affiliation{Hochfeld-Magnetlabor Dresden (HLD), Helmholtz-Zentrum Dresden-Rossendorf and TU Dresden, D-01314 Dresden, Germany}

\author{Y.-K. Huang}
\affiliation{Van der Waals-Zeeman Institute, University of Amsterdam, 1018XE Amsterdam, The Netherlands}

\author{J.A.~Mydosh}
\affiliation{Kamerlingh Onnes Laboratory and Institute Lorentz, Leiden University, 2300 RA Leiden, The Netherlands }

\date{\today}

\pacs{75.25.-j, 75.30.-m}
\begin{abstract}
We report the high-field induced magnetic phases and phase diagram of a high quality \urxrs~single crystal prepared using a modified Czochralski method. Our study, that combines high-field magnetization and electrical resistivity measurements, shows for fields applied along the $c$-axis direction three field-induced magnetic phase transitions at $\mu_{0} H_{c1}$ = 21.60 T, $\mu_{0} H_{c2}$ = 37.90 T and $\mu_{0} H_{c3}$ = 38.25 T, respectively. In agreement with a microscopic up-up-down arrangement  of the U magnetic moments the phase above $H_{c1}$ has a magnetization of about one third of the saturated value. In contrast  the phase between $H_{c2}$ and $H_{c3}$ has a magnetization that is a factor of  two lower than above the $H_{c3}$, where a polarized Fermi-liquid state with a saturated moment $M_{s}$ $\approx$  2.1 $\mu_{B}$/U is realized. Most of the respective transitions are reflected in the electrical resistivity as sudden drastic changes. Most notably, the phase between $H_{c1}$ and $H_{c2}$ exhibits substantially larger values. As the temperature increases, transitions smear out and disappear above $\approx$ 15 K. However, a substantial magnetoresistance is observed even at temperatures as high as 80 K. Due to a strong uniaxial magnetocrystalline anisotropy a very small field effect is observed for fields apllied perpendicular to the $c$-axis direction.

\end{abstract}

\maketitle

\section{Introduction}
Despite numerous studies, the exact nature of an order phase appearing in the heavy fermion compound \urs~below T$_{HO}$ = 17.5~K is still unclear. This phase, called hidden order (HO), is one of the most debated topics in heavy-fermion physics research\cite{*Mydosh11,Mydosh14}. Studies are often devoted to states emerging from the HO state. At low temperatures, starting from the HO state, different magnetic phases can be induced by pressure, magnetic field or a moderate substitution. The determination of such phases is a subject of many current studies \cite{Williams,Williams17,Shirer17,Prokes17a,Wilson}.

This HO state coexists below T$_{sc}$ = 1.5~K with superconductivity (SC) and is linked to a parasitic antiferromagnetic (AF) \cite{Palstra85,Hasselbach,Broholm} order characterized by a propagation vector \textbf{\textit{Q$_ {0}$}} = (1 0 0) with very small dipolar magnetic moments (0.01 - 0.03 $\mu_{B}$ \cite{Broholm}). These moments that seem to be related to lattice imperfections or strain are not compatible with a large entropy and a lambda-type specific heat anomaly at T$_{HO}$ associated with this transition. Consequently, the phase is called the small moment antiferromagnetic (SMAF) phase. Single crystals of higher quality with a higher residual electrical resistivity show as a rule smaller dipolar U moments. These moments, however, can develop either under pressure or with light doping of different dopants (Re, Rh, Fe, Os, Tc)  \cite{Burlet,Yanagisawa,Wilson,Williams,Butch,Baek,Das}. A common effect is that the HO and SC states are suppressed fairly quickly and new types of a magnetic order appear. In particular, Rh for Ru at a level of $\approx$ 2 \% stabilizes the SMAF and the HO re-appears at elevated fields but $\approx$ 4 \% Rh destroys the HO completely. Short-range AF correlations start to develop around 5 \% of Rh doping and lead to AF order at higher concentrations \cite{Burlet}. Above $\approx$ 10 \%, a long-range AF order with \textbf{\textit{Q$_ {3}$}} = ($\frac{1}{2}$ $\frac{1}{2}$ $\frac{1}{2}$) exists. By studying these emergent phases one hopes to be able to deduce valuable information on how the HO relates to the formation of U magnetic moments. 

While a moderate pressure stabilizes a long-range AF order of the \textbf{\textit{Q$_ {0}$}} type \cite{Amitsuka1999} (called large moment AF, LMAF), the applied magnetic field induces yet new phases. Polarized Fermi-liquid state with large saturated moments can be reached above a critical field  $\mu_{0} H_{c}$ $\approx$  38 T that seems to be independent on the Rh-doping level \cite{Kim03,Jaime,Oh,Knafo}. This phase is reached via a series of transitions that, in contrast, depends significantly on the Rh content. Concominant with a destruction of the HO state, the first critical field shifts to lower values \cite{Burlet,Sakakibara1992,Kuwahara,Prokes17b}.  Challenging high-field neutron diffraction experiments showed that in the case of the pristine system the first field induced phase adopts propagation vector \textbf{\textit{Q'$_ {2}$}} =  (0.60 0 0), close to the Fermi level nesting vector \cite{Knafo}. This structure is reported to be sine-wave modulated and perhaps of a multi-$Q$ nature. In contrast, for Rh-doped systems a commensurate up-up-down magnetic structure with propagation vector \textbf{\textit{Q$_ {2}$}} = ($\frac{2}{3}$ 0 0) has been detected \cite{Kuwahara,Prokes17b}.

Recently we have shown that \urxrs~ does not exhibit any sign of HO, SC, SMAF or LMAF states down to 0.2 K \cite{Prokes17a}. A heavy-fermion behavior, however, remains intact resembling above T$_{HO}$ very strongly properties of the pristine \urs~\cite{Prokes17a,Mydosh17}. Nevertheless, in contrast to \urs, it exhibits a short-range order (SRO) at \textbf{\textit{Q$_ {3}$}} \cite{Prokes18a}.  The SRO signal appears in zero field at temperatures comparable to T$_{HO}$ and it is initially strengthened by the applied field lower than $\approx$ 22 T where the first metamagnetic transition (MT) occurs. Above this field the SRO signal disappears, being replaced by new Bragg reflections indexable by \textbf{\textit{Q$_ {2}$}} \cite{Kuwahara,Prokes17b}. 

Up to date high-field experiments on \urxrs~above ordinary laboratory fields were limited to a single magnetization curve up to ~55 T recorded at  1.4 K and to neutron diffraction up to 24 T. In this work we report the high-field magnetic phase diagram up to 58 T deduced from combined measurements of the magnetization and electrical resistivities. We show that the response of the material to the external magnetic field is very anisotropic with the $c$-axis being the easy magnetization direction along which three MT's are detected. Sharp transitions indicate field-induced phases of a different kind, the first being an uncompensated AF with magnetization $\approx$ $M_{s}$/3, the second exhibiting magnetization of about $\approx$ $M_{s}$/2 before entering the fully polarized state. However, the intermediate $M_{s}$/2 phase is not clearly resolved in the electrical transport properties. 

Combining the available data the magnetic phase diagram is constructed documenting existence ranges of several magnetic phases and regimes of \urxrs. Results are discussed in a context of pristine and lightly doped \urs~systems.

\section{Experimental}

The details of our \urxrs~single crystal preparation, quality and other physical properties are presented in our recent work \cite{Prokes17a}. Magnetization $M$($T$) measurements in fields up to 58~T generated by discharging a capacitor bank producing a 25~ms long pulse were performed at the Hochfeld-Magnetlabor Dresden (HLD), Helmholtz-Zentrum Dresden-Rossendorf. A small 12.5 mg single crystal used in the magnetization experiment has been oriented using LAUE X-ray backscatter method. The edges of the sample were cut along directions parallel to principal crystallographic axes using spark erosion. Measurements were carried out between 1.4 K and 80 K with the field oriented along and perpendicular to the $c$-axis. The magnetic signal was detected using compensated pick-up coils and and scaled to match previous magnetization measurements obtained in static fields up to 14 T \cite{Prokes17a}. 

Electrical resistivity measurements were performed between 1.4 and 150 K on two bar-shaped single crystals (one of them used also for the magnetization measurements) with dimensions of about 0.7 mm x 0.7 mm x 3 mm cut along the principal $a$-and $c$-axes. We have used a standard four-wire AC method with frequencies between 16 and 25 kHz and excitation currents between 10 and 20 mA flowing along the longest dimension. The field has been applied along and perpendicular to the $c$-axis and both, longitudinal and transverse resistivities were recorded. Fields up to 62 T produced by a magnet with the pulse length of 150 ms were applied. 

\section{Results}

\subsection{Magnetization}

\begin{figure}
\includegraphics*[scale=0.30]{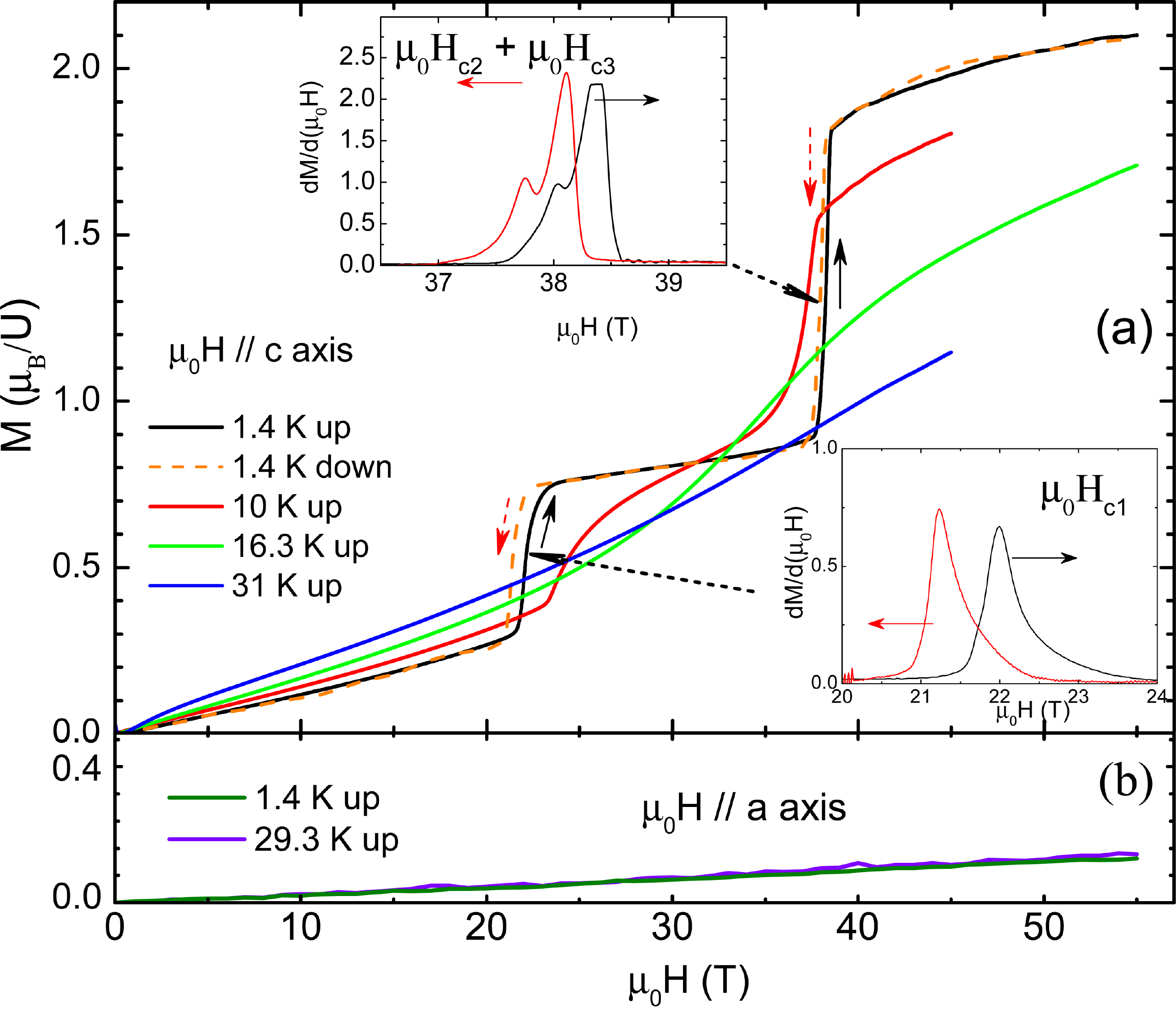}
\caption{(Color online) Field dependence of the \urxrs~single crystal magnetization measured at several different temperatures with pulsed fields applied along the $c$-axis (a) and perpendicular to the tetragonal axis (b). Insets in (a) show the field dependence of the field derivation of the magnetization at 1.4 K around metamagnetic transitions taken with increasing and decreasing field applied along the $c$-axis, respectively.} 
\label{fig1}
\end{figure}

The magnetic response of \urxrs~to the magnetic field is extremely anisotropic. In Fig.~\ref{fig1} (b) we show magnetization curves obtained at 1.4 K and at 29.3 K with field applied along the $a$-axis. The magnetization is small and increases linearly as a function of the applied field. No anomaly is present up to 58 T. Magnetic curves are very weakly temperature dependent, in agreement with low-field PPMS data. \cite{Prokes17a}

In contrast to the direction perpendicular to the tetragonal axis, magnetization measured with field along the $c$-axis exhibits strong field and temperature dependences. In Fig.~\ref{fig1} (a) we show several representative magnetization curves measured with field applied along the $c$-axis between 1.4 and 31 K with increasing field. For the lowest temperature we show also the descending field magnetization curve. At low temperatures, steep changes around $\approx$ 22 T and around $\approx$ 38 T, respectively, are observed marking metamagnetic transitions (MTs). The first MT, from the low-field heavy fermion liquid (HFL) to the first field-induced phase appears with increasing field at 22.0 T. As it is shown in insets of Fig.~\ref{fig1} (a), all the MTs appear with decreasing field at slightly lower fields. The first MT appears with decreasing field at 21.2 T leading to a hysteresis of $\approx$ 0.8 T and the average of field-up and down transitions at $\mu_{0} H_{c1}$ = 21.6 T. The significant hysteretic behavior is in agreement with first-order type of the phase transition found in the pristine system and 4 \% Rh-doped system transition \cite{Jaime,SilhanekPB,SilhanekPRL}.

A closer look at the transition around 38 T reveals that it consists from two anomalies suggesting a presence of two individual MT's. As it is illustrated in the upper inset of Fig.~\ref{fig1} (a), the former MT appears with increasing field at 38.05 T and the latter at 38.40~T, respectively. On the decreasing direction, they are found at 37.75 T and 38.10 T, respectively suggesting a somewhat smaller hysteresis of 0.3-0.4 T. The average of up and down sweeps amount to $\mu_{0} H_{c2}$ = 37.90 T and $\mu_{0} H_{c3}$ = 38.25 T. 

Above H$_{c3}$, the magnetization exhibits a gradual tendency towards a saturation at a level of M$_{s}$ = 2.1 $\mu_{B}$/U, which appears to be larger than in the pure and 4 \% Rh-doped systems.\cite{Kuwahara} Here a Fermi-liquid polarized state is established. The magnetization step across the first MT amounts to 0.46 $\mu_{B}$/U and across the second MT at 38 T to 0.94 $\mu_{B}$/U leading to a total magnetization change across MTs of 1.40 $\mu_{B}$/U. The increase at the former MT amounts thereby to one third of the total magnetization increase across all the transitions. This value is in accord with our recent high-field single crystal neutron diffraction on this system showing that the first field-induced phase is a commensurate uncompensated AF phase of 1.45 (9) $\mu_{B}$ U moments directed along the $c$-axis. The U moments are arranged in an up-up-down sequence propagating along the $a$-axis \cite{Prokes17b}. We denote therefore this phase as $M_{s}$/3 phase. On the contrary, compared to both pristine system and lightly Rh-doped systems, this phase exists over much larger range of fields between $\mu_{0} H_{c1}$ = 21.6 T and $\mu_{0} H_{c2}$ = 37.90 T \cite{Kim2004}. 

The magnetization between $H_{c2}$  and $H_{c3}$ amounts to about one half of the magnetization increase due to all MTs. We denote this phase as $M_{s}$/2 phase. The range of fields where this phase exists is very small, about 0.35 T. This is to be compared with larger ranges of existence of different field-induced phases in the pure and lightly Rh-doped systems that were studied in high-field magnetic fields \cite{Jaime,Kim2004,Kuwahara,Oh,Knafo,Prokes17b}.

The high-field magnetic susceptibility defined as $\chi (H)$~=~$\partial M(H)$/$\partial H$, where $H$ denotes the field strength are at 1.4 K distinctly different for different field-induced phases. While below $\mu_{0} H_{c1}$ = 21.6 T the $\chi (H)$ increases progressively in the vicinity of the transition, it is constant between $H_{c1}$ and $H_{c2}$ on a lower level than below $H_{c1}$. Above $H_{c3}$, the $\chi (H)$ at 1.4 K steadily decreases with increasing field towards a gradual saturation. 
    
With increasing temperature all MTs along the $c$-axis direction smear and the hysteresis between increasing and decreasing field branches becomes reduced. The first MT, defined by the maximum on the $\partial M(H)$/$\partial H$ as function of $H$, moves with increasing temperature steadily to higher fields and disappears above $\approx$ 15-16 K. In contrast, the upper two MT's shift with increasing temperature to lower fields. Although all the transitions are relatively well defined we could follow the splitting of the transitions in the field derivatives of $M(H)$ around 38 T only up to $\approx$ 9 K. Above this temperature we observe around 38 T a single MT. 
 
Between $\approx$ 9 K and $\approx$ 16 K, well defined maxima in $\partial M(H)$/$\partial H$ denote the field range where the $M_{s}$/3 phase exists. Between these maxima the $M$($H$) exhibits an S-shape type increase (see Fig.~\ref{fig1}). Above $\approx$ 16 K up to $\approx$ 50 K no clear anomalies are visible from the $M$($H$) dependences. However, one still observes an S-shaped magnetization curve. We interpret the inflection point as a field above which a field-induced polarized state exists in \urxrs. At yet higher temperatures the $M$($H$) increases linearly with field.

\begin{figure}
\includegraphics*[scale=0.22]{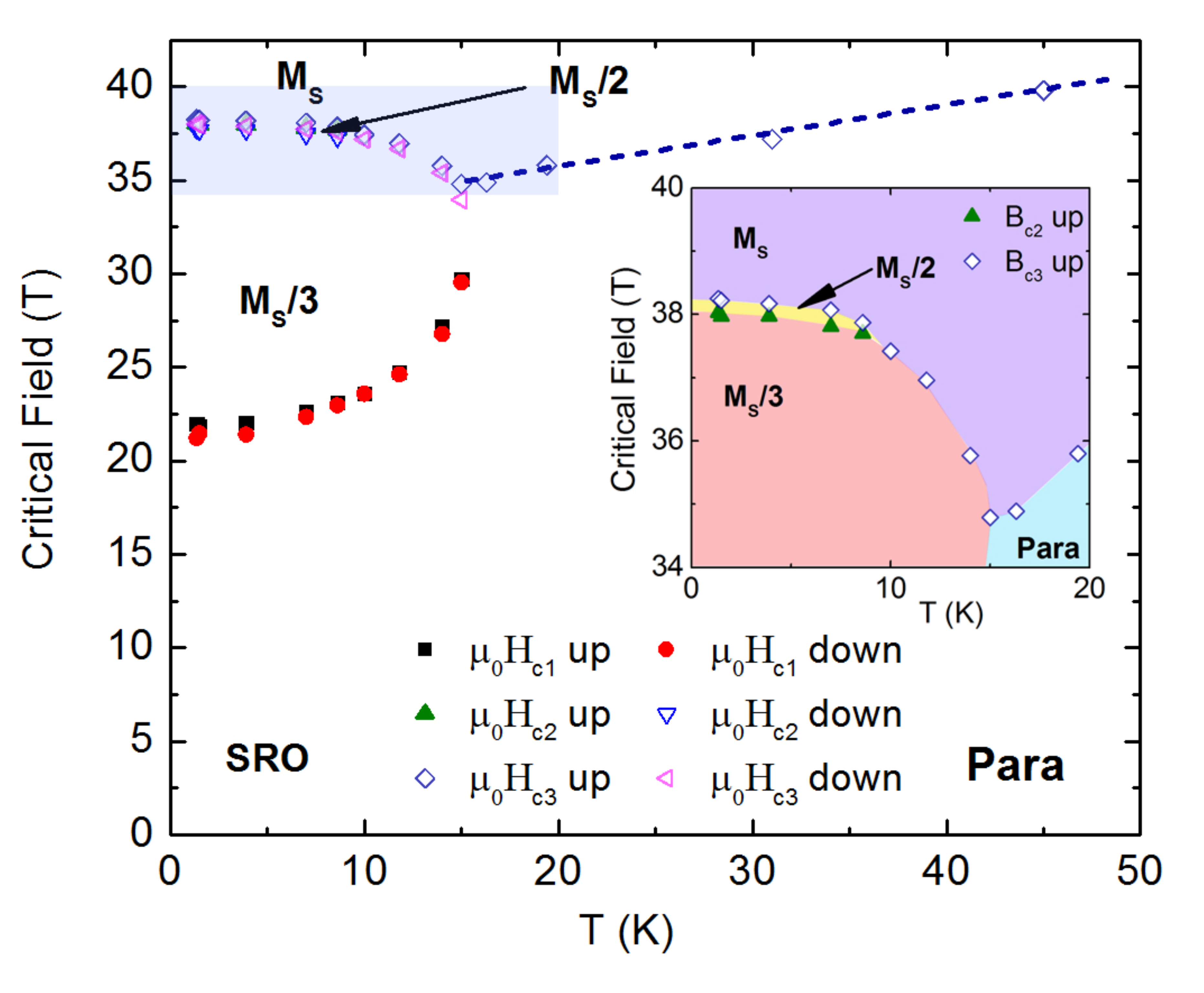}
\caption{(Color online) Magnetic phase diagram of \urxrs~single crystal  deduced from magnetization measurements with field applied along the tetragonal axis. Para, SRO, M$_{s}$, M$_{s}$/2 and M$_{s}$/3 denote various phases existing in \urxrs~(see the main text).} \label{fig2}
\end{figure}

In Fig.~\ref{fig2} the magnetic phase diagram for field applied along the tetragonal axis is constructed from high-field magnetization data. In the inset we show the detail of the range where the $M_{s}$/2 phase exists and can be traced up to up to $\approx$ 9 K. The magnetic phase diagram also clearly establishes that the $M_{s}$/3 phase occurs as an inclusion between the low-field state where only the short-range order (SRO) defined by \textbf{\textit{Q$_ {3}$}} exists and the $M_{s}$/2 phase, or, at temperatures between $\approx$ 9 K and $\approx$ 15 K, between the SRO and the polarized state.

\subsection{Electrical Resistivity}
    
Electrical resistivity measurements with field applied along the $a$-axis show no anomalies (not shown). This finding is in agreement with magnetization data which show no appreciable field effect for this orientation. The observed change of the resistivity with current both along the tetragonal axis and perpendicular to it is qualitatively similar to data published for the pure system \cite{Scheerer12,Scheerer14}. In the pure system, however, larger changes at low temperatures are found. In our sample, for the current along the $c$ axis, the resistivity increases by less than $\approx$ 4 \% of its zero-field value, indicating that the field applied perpendicular to $c$-axis direction does not alternate significantly the scattering of conduction electrons. 

\begin{figure}
\includegraphics*[scale=0.18]{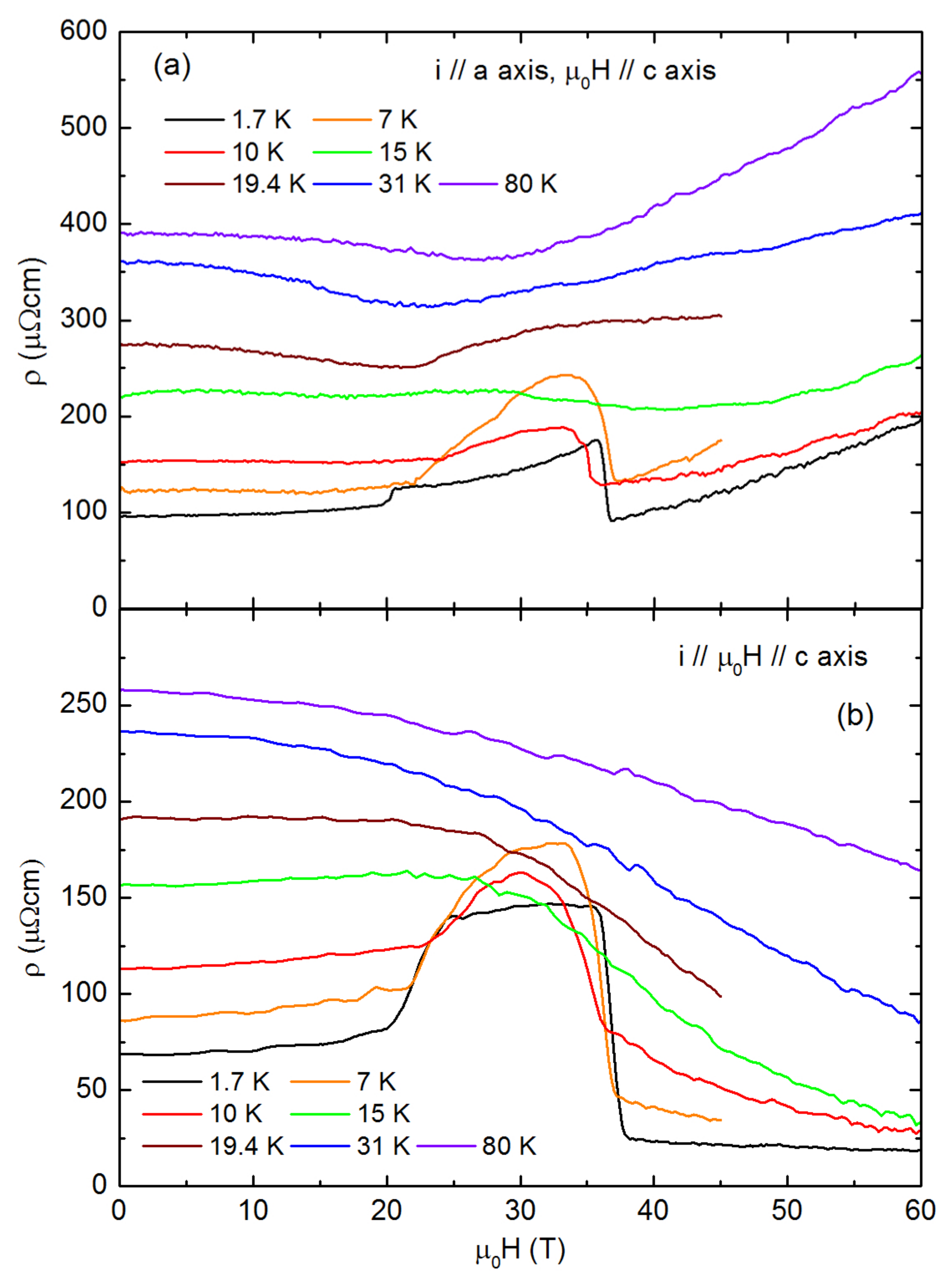}
\caption{(Color online) Field dependence of the \urxrs~single crystal electrical resistivity measured at several different temperatures with pulsed fields (sweeps down) applied along the $c$-axis with electrical current perpendicular (a) and along (b) the tetragonal axis.} 
\label{fig3}
\end{figure}

In Fig.~\ref{fig3} (a) the field dependence of the electrical resistivity with current along the $a$-axis, $\rho_{a}(H//c)$, and field applied along the tetragonal axis for selected temperatures between 1.7 K and 80 K is shown. These data are taken with decreasing field. At 1.5 K, the first MT at $H_{c1}$ manifests itself as a sudden increase of the resistivity. The transition to the polarized state causes, in contrary, a significant decrease. We are not able to resolve the two transitions at $H_{c2}$ and $H_{c3}$ that were indicated in the magnetization measurements. As the temperature increases, both MTs smear out, the one at lower field faster than the upper one. 

 It should be noted that at the lowest temperature, below the first MT and between MTs the resistivity cannot be approximated by the expression $\rho_{H}$ = $\rho_{0 T}$ + $a H^{2}$, where $\rho_{0 T}$ denotes the electrical resistivity at zero field. To get a reasonable description of data, an inclusion of a term linear in $H$ is necessary documenting beyond Fermi liquid type behavior. This type of field dependence that is in contrast to the pure system where the $H^{2}$ behavior is observed \cite{Palstra85,Uwatoko92}. Such behavior has been previously identified for the low-field range at various temperatures for this sample. The agreement of the fitted parameters with literature is good \cite{Prokes17b}. Above the  $H_{c3}$,  the electrical resistivity depends quadratically on the magnetic field. This type of fit yields  $\rho_{0 T}$ = 25.8 (5) $\mu\Omega$cm and suggests that much of the electron scattering is caused by processes that are quenched in the high field limit.

The field dependence of the resistivity changes with temperature significantly. For instance, the behavior below the first MT is at high temperatures reverted to that at low temperatures and the resistivity decreases with increasing field. Up to $\approx$ 15 K the measured curves have a dome-like dependence with higher resistivities between $H_{c2}$  and $H_{c3}$. At higher temperatures one can discern a reduction of the electrical resistivity values up to 20-30 T followed by a subsequent increase at higher fields. At the highest temperatures the increase in the high-field limit is nearly linear.

In Fig.~\ref{fig3} (b) we show the field dependence of the electrical resistivity $\rho_{c}(H//c)$ with field and current along the $c$-axis. Also these data are taken with descending fields. At 1.7 K the MTs are clearly visible and manifest themselves as sudden changes in the electrical resistivity values. Starting from the relatively high resistivity at zero field, the resistivity increases moderately in agreement with PPMS data \cite{Prokes17a}, first by few percents until $\approx$ 20 T where it starts to increase significantly up to $\approx$ 24 T. Here it attains twice as high resistivity with respect to the zero field. The interval across which it increases is significantly broader than the transition indicated in magnetic bulk measurements. However, it still coincides with the the first MT centered at $\mu_{0} H_{c1}$ = 21.6 T. In the M$_{s}$/3 phase the resistivity increases nearly linearly by very few \% only to drop above $\approx$ 36 T. The reduction over a field interval of $\approx$ 2 T across this transition is significant and the resistivity reaches to about 1/3 of its zero field value with a further weak decrease at even higher fields. At 1.7 K, the electrical resistivity reduces at 60 T by 74 \% with respect to its zero field value. Extrapolation of data taken at the lowest temperature assuming a linear dependence towards zero field suggests $\rho_{0 T}$ = 32 (1) $\mu\Omega$cm, i.e. a value comparable with the extrapolation of the data at 1.7 K for electrical current along the $a$-axis. 

While a hysteresis of a comparable width as in the magnetization measurements has been detected across $H_{c1}$ in electrical resistivity, no significant hysteresis between field up and field down sweeps has been detected in the $H_{c2}$-$H_{c3}$ region. However, the critical field of MTs defined from electrical resistivity for both orientations are found at different fields. Also the width of the transition is different - the electrical transport yield broader transitions.

\begin{figure}
\includegraphics*[scale=0.28]{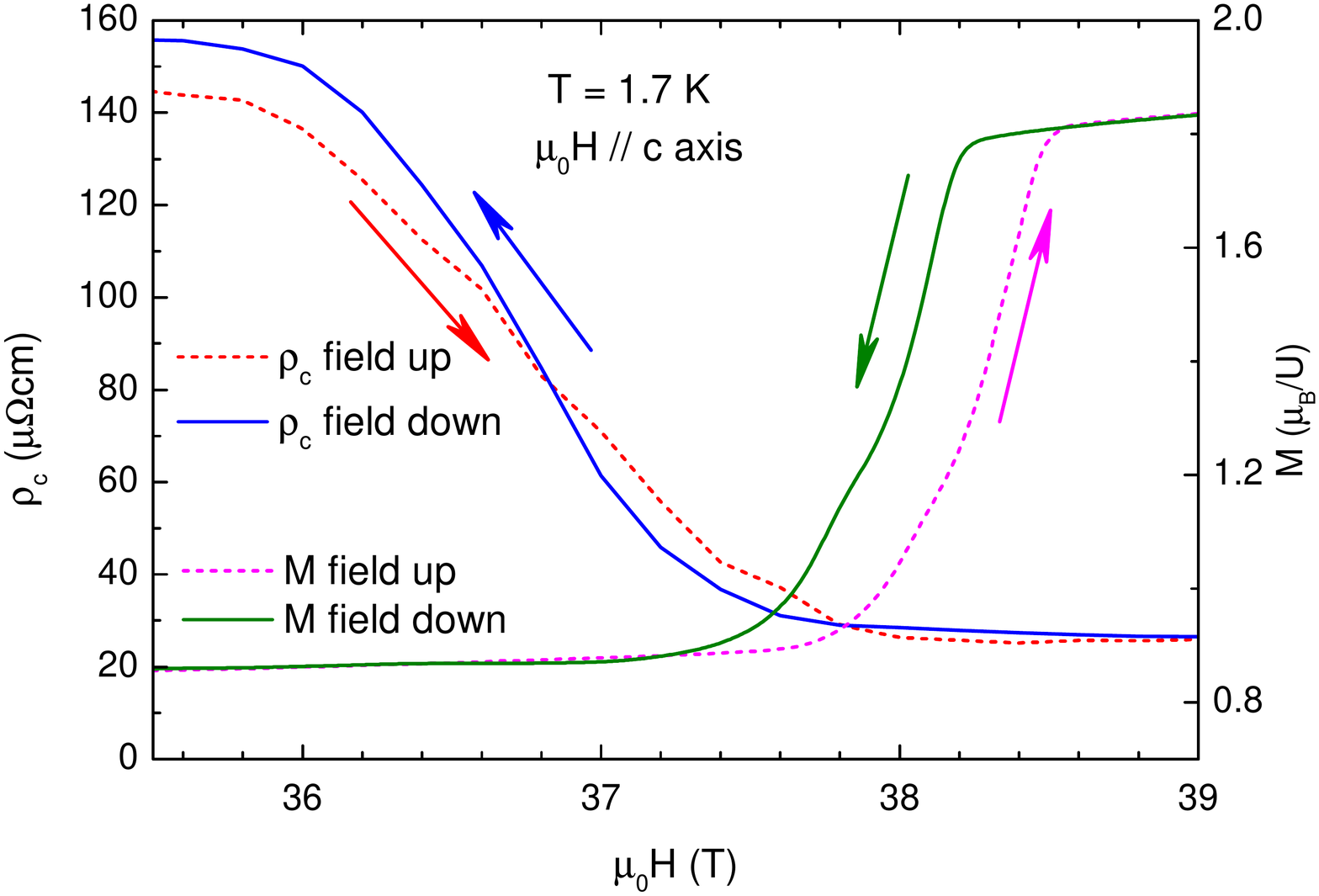}
\caption{(Color online) Enlarged portion of the field dependence around the $H_{c2}$ - $H_{c3}$ transitions of the $\rho_{c}(H)$ for increasing (dashed line) and decreasing (full line) field sweeps as compared to the magnetization curve, measured with field applied along the $c$-axis. Note different onsets of relevant changes and width of transitions.} 
\label{fig4}
\end{figure}

As it is shown in Fig.~\ref{fig4}, with increasing field the electrical resistivity $\rho_{c}(H)$ reaches its reduced value at fields where the magnetization only starts to increase. Similarly, upon decreasing the field first the magnetization reaches a reduced value below $H_{c2}$ before the resistivity starts to increase. At the same time, the transition range seen on the electrical resistivity is about three times broader (10 \% - 90 \% rule) than the transition across both $H_{c2}$ - $H_{c3}$ transitions seen on the magnetization. This is valid for both, ascending and descending fields, ruling out a possibility that the difference is due to a different preferred way of data recording (the electrical resistivity presented above have been measured as a rule with descending field while the magnetization with increasing field). Similar observation, albeit with smaller differences between the magnetization and resistivity, is found for the first MT. The electrical resistivity starts to change before the relevant change starts on the magnetization (not shown).

With increasing temperature the zero field resistivity increases in agreement with previous data \cite{Prokes17a} and the changes at respective critical fields are smeared out. One observes a dome-like structure for temperatures up to $\approx$ 15 K. Similar to the $a$-axis direction, no clear indication of the M$_{s}$/2 phase s found.

Comparing the measurements with current along the $a$-and $c$-axis one realizes several significant differences. First, the electrical resistivity along the $a$-axis is larger than along the tetragonal axis at all temperatures except for a small field interval at lowest temperatures. Second, in the M$_{s}$/3 phase $\rho_{c}(H)$ increases at 1.7 K nearly linearly with field, changing at intermediate temperatures between $\approx$ 5 K and $\approx$ 13 K to a dome-like dependence similar to the $a$-axis direction. Finally, at high enough temperatures (above $\approx$ 20 K) the $\rho_{c}(H)$ decreases with field, in contrast to $\rho_{a}(H)$ that first decreases and then increases in the high field limit. This difference results from a different geometry of the current with respect to the applied field and is due to cyclotron motion of electrons that leads to higher scattering rate in the tranverse geometry.

            \begin{figure}
\includegraphics*[scale=0.32]{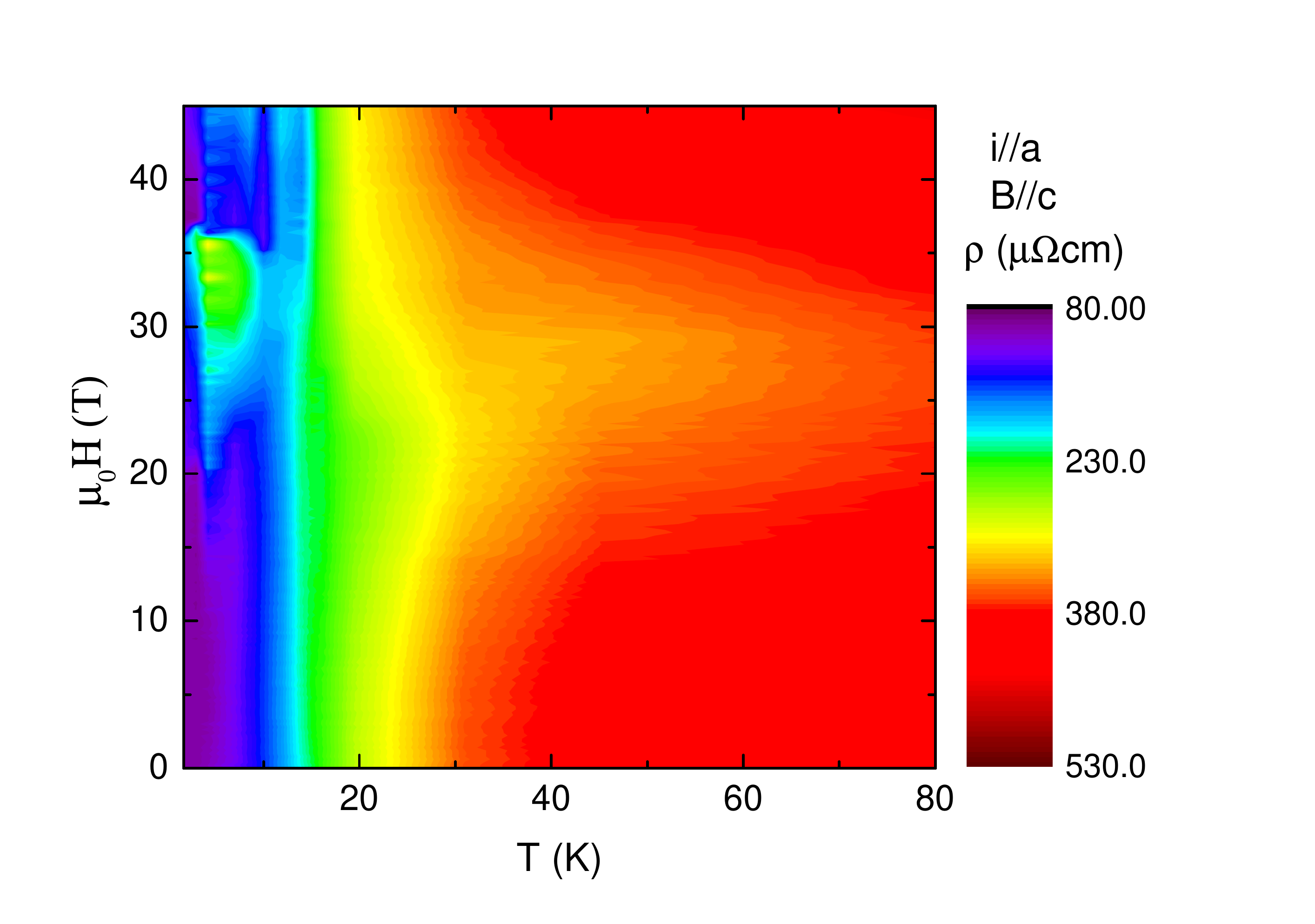}
\caption{(Color online) Color-coded transverse electrical resistivity $\rho_{a}(H)$ of the \urxrs~single crystal with current along the $a$-axis as a function of the temperature and magnetic field applied along the $c$-axis.} 
\label{fig5}
\end{figure}

            \begin{figure}
\includegraphics*[scale=0.32]{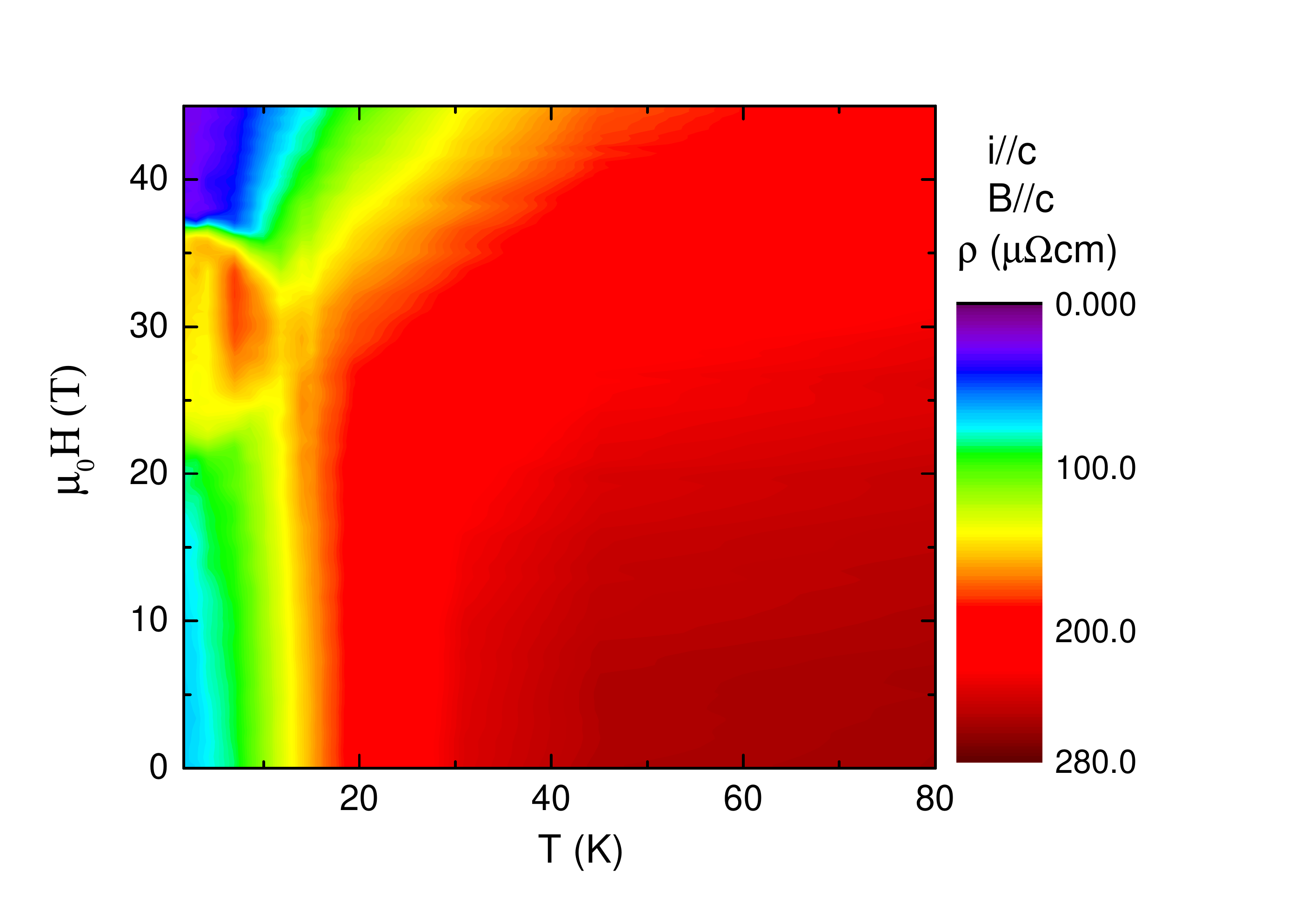}
\caption{(Color online) Color-coded longitudinal electrical resistivity $\rho_{c}(H)$ of the \urxrs~single crystal measured along the $c$-axis as a function of the temperature and magnetic field applied along the $c$-axis.} 
\label{fig6}
\end{figure}

A more comprehensive picture of the electrical resistivity behavior can be obtained form color-coded maps in Fig.~\ref{fig5} and  Fig.~\ref{fig6} that show a portion of the transverse $\rho_{a}(H)$ and longitudinal $\rho_{c}(H)$ electrical resistivities with current along the $a$-axis and the $c$-axis, respectively as a function of temperature and magnetic field applied along the $c$-axis. Clearly, the $M_{s}$/3 manifests itself as an island of enhanced resistivity for both current directions. Common for both current directions is a rather significant reduction of resistivity values at low temperatures that is clearly present  for both orientations also at high fields. This observation is similar to measurements on the pure system. An exception is the $c$-axis orientation at high fields above the $M_{s}$/3 phase, where the low-resistivity region extends as compared to the $a$-axis direction also to higher temperatures. Also common for both orientations is a decrease of the resistivity with increasing field up to $\approx$ 25 T at elevated temperatures, where the reduction is progressively larger in the low- temperature limit. Above this field only the $c$-axis resistivity continues to decrease. For the $a$-axis direction the resistivity increases in the high-field limit leading to a minimum around 25 T. The latter observation indicates different contributions to the scattering of conduction electrons. 
 
\section{Discussion}

Presented data show unambiguously that the response of \urxrs~to magnetic field is extremely anisotropic. This is documented by the field induced metamagnetic transitions for the field applied along the $c$-axis observed at $\mu_{0} H_{c1}$ = 21.60 T, $\mu_{0} H_{c2}$ = 37.90 T and $\mu_{0} H_{c3}$ = 38.25 T, respectively and the absence of any anomalies on the magnetization curve for the direction perpendicular to the tetragonal axis. 
This behavior resembles very much properties of the pure system. The magnetocrystalline anisotropy that has Ising character is caused by the non-Kramers doublet in $\Gamma_{5}$ that couples only to the $c$-axis component of the magnetic field. The two 5$f^{2}$ states, having both electric quadrupolar and spin degrees of freedom (spin-orbital liquid \cite{Schaffer}) are proposed to be responsible for the hidden order in the pristine system and the first-order phase transition at $H_{c1}$ \cite{Chandra,SilhanekPRL}. In our, 8 \% Rh-doped system, no HO exists. However, the bare physical properties remains the same - absence of a long-range magnetic order and heavy-fermion behavior at low temperatures, very strong uniaxial anisotropy and field-induced phases for field along the $c$-axis before arriving to a polarized state. It is therefore to be expected that in the low-T limit at low fields also in \urxrs~ are 5$f$ electrons in the 5$f^{2}$ configuration. These become itinerant upon application of strong magnetic field. A Zeeman splitting causes one of the subbands to be shifted below the Fermi energy causing the reconstruction of its topology, reducing the hybridization between the 5$f$ states and conduction electrons and thereby creating sizable and stable magnetic moments. These appear at low temperatures above $H_{c1}$. As the application of field perpendicular to the $c$-axis does not reconstruct the Fermi topology, the 5$f$ electrons are still  in the vicinity of the Fermi surface, hybridized with conduction electron.

This is also the reason why the anisotropy is reflected very strongly in the electrical transport properties. While for the field applied along the $a$-axis no significant modifications are observed in agreement with the magnetization behavior, the application of the field along the tetragonal axis leads to drastic changes of the resistivity along both the $a$-and $c$-axes. At low temperatures, the state between $H_{c1}$  and $H_{c2}$ leads generally to a larger resistivity than below the MT at $H_{c1}$ where merely a short-range order exists and in the high field limit where the polarized state is established. The simplest way to interpret this observation is that it is due to the appearance of a long-range magnetic order (phase M$_{s}$/3) with \textbf{\textit{Q$_ {2}$}} = ($\frac{2}{3}$ 0 0) leading to a modification of the Fermi surface topology and the appearance an additional superzone boundary. As the details of the entire Fermi surface topology influence conduction electron scattering (also due to strong hybridization with other electron states), it is not surprising that one observes signatures of the M$_{s}$/3 phase (and generally the influence of field applied along the $c$-axis) for both current orientations. The significant reduction of the electrical resistivity upon entering the polarized phase, called giant magnetoresistance, is very common in uniaxial U-based systems \cite{sech98,Prokes96} and is present also in the pure system \cite{Scheerer14,Scheerer12}. It is usually interpreted as being due to anisotropic reconstruction of the Fermi surface topology and a reduction of anisotropic magnetic fluctuations  \cite{sech98,sechovsky95}. 

Further in the high field limit, especially at elevated temperatures, the electrical resistivities along and perpendicular to the tetragonal axis behave differently. For the $c$-axis direction the resistivity values above $H_{c3}$ are at all temperatures significantly lower than in zero fields. In contrast, the resistivity for the $a$-axis direction, i.e. with the current perpendicular to the applied field, increases at high fields. This different behavior could be explained considering the combined effects of the magnetocrystalline anisotropy and the Lorentz force acting on the conduction electrons causing classical magnetoresistance effect. Higher cyclotron frequencies lead to reduction of the mean free path of conduction electrons and higher scattering rates for the transverse geometry. Comparison with available data for the pure system \cite{Scheerer14} we recognize that the effect of the magnetoresistance is stronger in our sample.

At this point we should mention a possible non-negligible influence of the magnetocaloric effect (MCE) and/or eddy currents. While the former effect can lead to either increase or decrease of the sample's temperature, the latter one always increases the sample temperature. MCE has been clearly identified in previous high field experiments in the pure and 4 \% Rh doped systems \cite{Jaime,SilhanekPRL,SilhanekPB}. This effect is rather important as it shifts the phase boundaries and obscures the exact determination of the existence regions of various phases. Also, it may hamper the comparison between results obtained using different techniques. In fact, our preliminary measurements using a much larger single crystal indicate that also \urxrs~exhibits MCE \cite{Prokes18b}. It is therefore almost sure that MCE plays a role also in our measurements and would lead to modifications of the phase transition values. However, we argue that even if the phase boundaries are determined with lesser precision, the main message of the current work remains intact. Question can be raised whether the double structure of the MT at $H_{c2}$ - $H_{c3}$ cannot be either due to variations in the Rh concentration or due to temperature inhomogeneities caused by fast field sweeps. Indeed, concentration inhomogeneities would lead to a broadening or even splitting of  MTs. However, no such structure is observed around $H_{c1}$ that is more sensitive to the Rh concentration. In addition, the MTs are sharp and neutron diffraction observation with resolution limited Bragg reflections did not reveal any structural inhomogeneities \cite{Prokes17a}. Therefore we conclude that the double MT at $H_{c2}$ - $H_{c3}$ is real. For analogical reasons we rule out also a possible influence of temperature inhomogeneities due to fast field sweeps. On the other hand, both mechanisms probably contribute to the observed hysteresis of all MTs.

In general, the electrical resistivity follows closely the field-induced changes of the \urxrs~magnetic state. Metamagnetic transitions at $H_{c1}$  and $H_{c2}$ are clearly visible both in magnetization and electrical resistivity. An exception here is the $H_{c3}$ transition that remains invisible in the electrical resistivity measurement. In the transport properties only $H_{c2}$ MT is visible, being completed before the magnetization changes (see Fig.~\ref{fig4}). As the resistivity experiments were performed with longer field pulses, eddy currents and MCE play less important role with the temperature of the sample being more constant. As eddy currents always increase the temperature of the sample, it is to expected that the sample during magnetization experiment (with higher field sweep rate) is higher than in the resistivity measurement case. Considering the magnetic phase diagram shown in Fig.~\ref{fig2}, this would lead to a lower critical fields of the upper transition(s) than determined from magnetization. However, it is the electrical resistivity that leads lower critical field around 38 T. This observation suggests that both, the polarized and the $M_{s}$/2 phases are created due to reconstruction of the Fermi surface leading to polarization of individual Fermi surface pockets \cite{Harrison13} and consequently to reduction of the 5$f$-conduction electron hybridization that short-cuts the sample just below $H_{c2}$.

            \begin{figure}
\includegraphics*[scale=0.32]{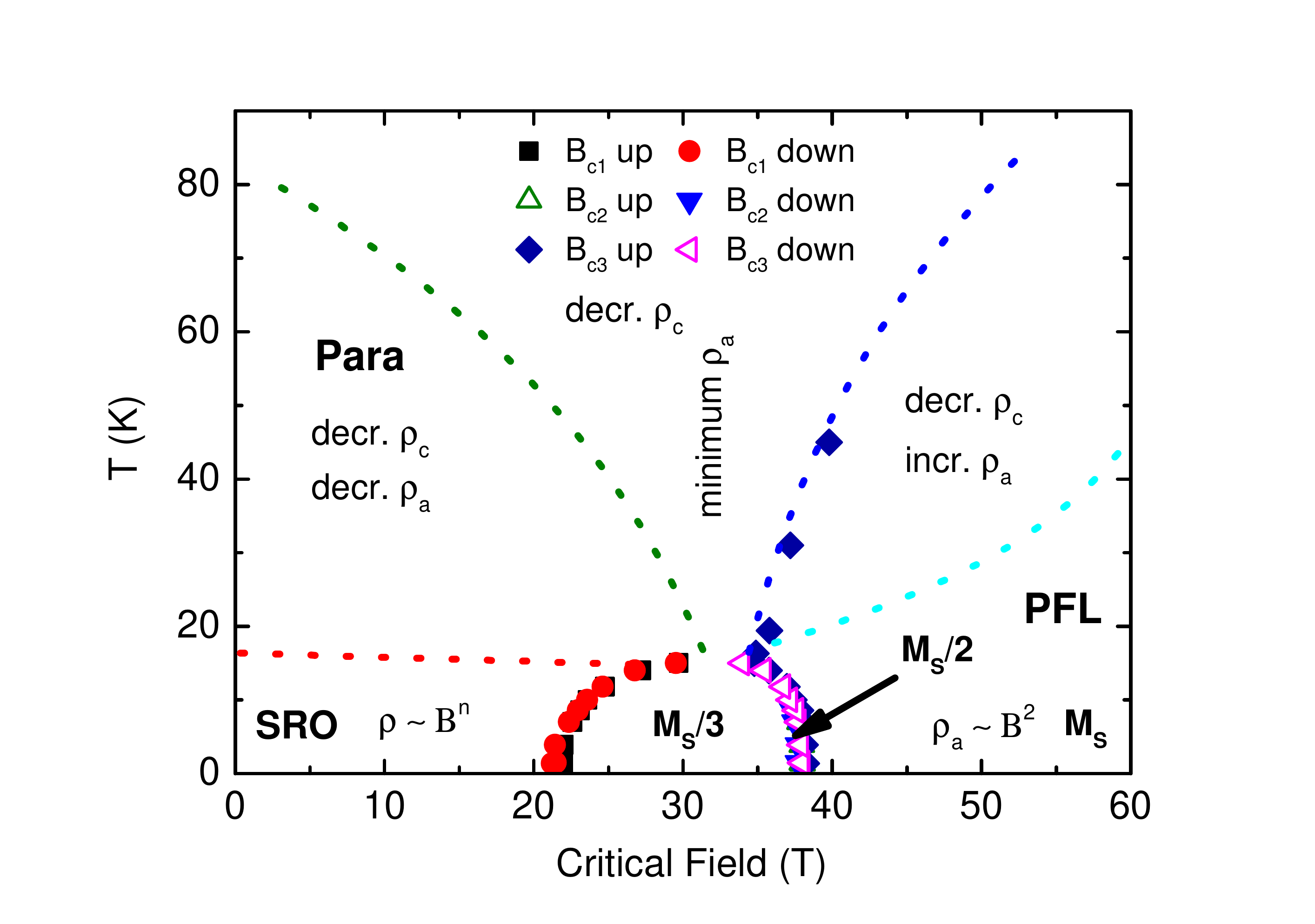}
\caption{(Color online) Magnetic phase diagram of \urxrs~for magnetic field applied along the $c$-axis documenting the ranges of existence of various phases and different regions with typical field and temperature behaviors of the electrical resistivity. SRO denotes region with a short-range magnetic order described by \textbf{\textit{Q$_ {3}$}}, PFL denotes Polarized Fermi liquid state, Para a paramagnetric state and $M_{s}$/3 and $M_{s}$/2 field induced phases with one third and one half saturated magnetization $M_{s}$, respectively. Dashed lines represent approximate boarders of different resistivity behavior regions (see the main text).} 
\label{fig7}
\end{figure}

From the available data we have constructed the magnetic phase diagram for the field applied along the $c$-axis that is shown in Fig.~\ref{fig7}. Besides the phase transition boundaries determined from the field and temperature anomalies of the magnetization and electrical resistivity we identify also regions of different electrical resistivity behavior as a function of temperature or the applied field. First of all, by the red dashed  line we denote position of the inflection point in the temperature dependence of the electrical resistivity that agrees with the occurrence of the SRO \cite{Prokes18a}. It is interesting to note that the SRO appears approximately at the same temperature as the HO in the pristine system. This may suggest that both ground states are on the same energy scale. Both ground states are itinerant heavy-fermion liquids. The difference is, however, a lesser degree of coherence of the SRO state and thereby magnetic disorder that is reflected also by much higher residual resistivity in Rh doped systems. The loss of coherence is most probably also the reason for destruction of the HO. The light blue dashed line limits the quadratic field dependence of the electrical resistivity found for current along the $a$-axis due to classical magnetoresistance. This type of dependence changes at higher temperatures to a nearly linear one. The green line shows approximately the high-temperature low-field region where the electrical resistivity along both $a$-axis and $c$-axis directions decreases suggesting a reduction of conduction electron scattering due to magnetic fluctuations. Finally, the dark blue line defines a region where the $a$-axis resistivity alone increases leading to a minimum between the two regions. Comparing our magnetic phase diagram with those published for the pristine and lightly doped Rh systems \cite{*Mydosh11,Mydosh14,Kim2004,Kim03,Jaime,Oh} one realizes immediately that they are very similar, especially in the high field region. Biggest differences concern the low-field region at low temperatures. In the pristine system, where the HO exists the longitudinal resistivity strongly increases to fall down upon entering the first induced phase. In our system there is no HO and also the electrical resistivity does not show any anomalies below $H_{c1}$ that is in addition shifted significantly to lower field values. This points to a significant interplay between conduction electron states and the HO order which is removed when replaced by SRO in our system. Yet, the SRO and the HO sets in at similar temperatures. Interesting is also the fact that the uppermost MT appears in the pristine and Rh-doped systems nearly at the same critical fields suggesting yet another common energy scale in these systems.

In conclusion, we show that the behavior of \urxrs~is very anisotropic and the electrical transport properties are dictated by the Fermi surface topology changes caused by the magnetic field. The easy magnetization direction is found along the tetragonal axis, similar to other systems stemming from the pure \urs. We identify several distinct field induced magnetic phases and ranges with different field/temperature behavior of the electrical resistivity pointing to various different scattering mechanisms of conduction electrons. Complementary information should become available from the Hall effect experiments.

\begin{acknowledgements}
 We acknowledge the support of the HLD at HZDR, member of the European Magnetic Field Laboratory (EMFL). 
\end{acknowledgements}

%

\end{document}